\documentclass[prl,twocolumn,showpacs,groupedaddress]{revtex4}
\usepackage[]{graphicx}

\begin{document}

\renewcommand{\baselinestretch}{0.95}

\title{Saturable discrete vector solitons in one-dimensional photonic lattices}

\author{Rodrigo A. Vicencio}\email{rodrigov@fisica.ciencias.uchile.cl}
\homepage{https://fisica.ciencias.uchile.cl/~rodrigov}

\affiliation{Max Planck Institute of Physics of Complex Systems,
01187 Dresden, Germany,} \affiliation{Departamento de F\'isica,
Facultad de Ciencias, Universidad de Chile, Santiago, Chile}

\author{Eugene Smirnov, Christian E. R\"uter, and Detlef Kip}
\affiliation{Institute of Physics and Physical Technologies,
Clausthal University of Technology, 38678 Clausthal-Zellerfeld,
Germany}

\author{Milutin Stepi\'c}
\affiliation{Department of Optical Technologies, National
Institute of Metrology, 38116 Braunschweig, Germany,}
\affiliation{Vin\v ca Institute of Nuclear Sciences, P.O.B. 522,
11001 Belgrade, Serbia}

\begin{abstract}
Localized vectorial modes, with equal frequencies and mutually
orthogonal polarizations, are investigated both analytically and
experimentally in a one-dimensional photonic lattice with
saturable nonlinearity. It is shown that these modes may span over
many lattice elements and that energy transfer among the two
components is both phase and intensity dependent. The transverse
electrically polarized mode exhibits a single-hump structure and
spreads in cascades in saturation, while the transverse
magnetically polarized mode exhibits splitting into a two-hump
structure. Experimentally such discrete vector solitons are
observed in lithium niobate lattices for both coherent and
mutually incoherent excitations.
\end{abstract}


\maketitle

Solitons or stable strongly localized nonlinear structures, which
can elastically interact with linear waves and other solitons,
have been studied in various systems in nature, ranging from
astrophysics \cite{1} and ocean waves \cite{2}, down to Josephson
junctions \cite{3} and nanowires \cite{4}. These localized
structures exist due to an exact balance between two or more
counteracting effects such as, for example, dispersion and
nonlinearity in the temporal domain \cite{book}. In the optical
domain, solitons may exist in specific materials, such as Kerr and
photorefractive ones \cite{6,7}. On the other hand, solitons occur
in different forms like incoherent, discrete, and vector solitons,
which are not directly related to a particular material
\cite{book}. Vector solitons are composite structures that consist
of two or more components which mutually self-trap in a nonlinear
medium. Importantly, the individual components decay in isolation.
The existence of vector solitons was first suggested by Manakov in
1974 \cite{9}. Later on, vector solitons have been, for example,
studied in carbon disulfide \cite {10} and photorefractive
crystals \cite{11}. In periodic nonlinear systems, the so-called
discrete solitons exist due to the balance between nonlinearity
and discreteness \cite{DS}. They have been observed in diverse
physical configurations such as biological systems,
charge-transfer solids, Josephson junctions, micromechanical
oscillator arrays, and photonic lattices. However, nonlinear
optics has overtaken a primacy in the nowadays soliton related
research \cite{15,16} due to a rather mature technology of
photonic lattice fabrication. Such photonic crystals have periodic
distributions of the refractive index and light propagation is
associated with allowed bands and forbidden gaps, analog to
propagation of electrons in crystalline lattices \cite{OE05}.

One-dimensional (1D) discrete vector solitons (DVS) originating
from the first band have been already investigated both
analytically \cite{17,18} and experimentally \cite{19} in
nonlinear cubic waveguide arrays (WA). The two-dimensional case
was studied, too \cite{20}. Finally, multi-band vector solitons,
in which individual components stem from different bands, were
also recently suggested and demonstrated \cite{21}. The aim of the
present study is to investigate DVS in media with saturable
nonlinearity. Prime examples of such media are photonic lattices
in photorefractive crystals. Additionally, semiconductors at higher light intensities also exhibit saturation \cite{22}. It has been shown that saturation, which may occur in a cascade manner in discrete systems \cite{23}, is responsible for the
existence of multiple zeros of the Peierls-Nabarro potential,
leading to free steering of large amplitude solitons, and stable
propagation of inter-site modes in 1D and 2D systems \cite{24}.
Various species of two-component saturable DVS have been
investigated recently, where it was assumed that both components
have the same polarization and different frequencies \cite{25}. In
what follows we are interested in the situation where the
components have the same frequency but differ in polarization.

By following the procedures outlined in Refs.~\cite{11,26},
assuming only nearest neighbor interactions, and by using the slowly
varying envelope approximation, one may obtain the following model
equations:
\begin{eqnarray}
&&i\frac{\partial u_n}{\partial \xi}+L_u u_n-\beta\ \frac{\left(u_n+s_2 B v_n\right)}{\left(1+|u_n|^2+s_1|v_n|^2\right)}=0,\nonumber\\
&&i\frac{\partial v_n}{\partial \xi}+L_v v_n-\,\beta\ \frac{\left(s_1 B u_n+s_3Av_n\right)}{\left(1+|u_n|^2+s_1|v_n|^2\right)}=0, \label{eqf}
\end{eqnarray}
where $L_u u_n= (C_0-\Delta k) u_n+V_0(u_{n+1}+u_{n-1})$ and $L_v
v_n=[C_0 v_n+V_0(v_{n+1}+v_{n-1})]/s_1$. Here $\xi$ is the
normalized propagation coordinate ($y$ in the experiment). The
normalized envelopes $u_n$ and $v_n$ correspond to transverse
electrically (TE) and magnetically (TM) polarized fields,
respectively. The parameter $\beta$ represents the nonlinear
coefficient, the normalized coupling constant is denoted by $V_0$,
$\Delta k$ is the normalized difference of TE and TM wave numbers,
whereas $C_0$ can be regarded as a normalized propagation
constant. By defining birefringence $\Delta n= n_x-n_z$ and
average refractive index $n_0=(n_x+n_z)/2$, we write the function
$s_j\approx 1+j\Delta n/{n_0}$. Finally, $A=r_{xxz}/r_{zzz}$ and
$B=r_{xzx}/r_{zzz}$, where $r_{ijk}$ denote the components of the
Pockels tensor \cite{11}. One may notice that in general, alike
DVS in Kerr media \cite{18,19,20}, in our situation there exists
no possibility to separate cross-phase and four-wave mixing
effects. A conserved quantity of this model is the total power,
$P=\sum_n\left(|u_{n}|^2+s_1|v_{n}|^2\right)=P_u+s_1 P_v$. By
using (\ref{eqf}), it can be shown that ${\partial P}/{\partial
\xi}=0$ implies:
\begin{equation}
\frac{\partial P_u}{\partial \xi}=-s_1 \frac{\partial
P_v}{\partial \xi}=2\beta s_2 B \sum_n \frac{\mathrm{Im} (v_n
u_n^*)}{\left(1+|u_n|^2+s_1|v_n|^2\right)}\,. \label{Pes}
\end{equation}
This expression gives us information on the energy (power)
exchange among the TE and TM components which clearly will depend
on the total level  of power in each waveguide. By considering a
one-channel constant-amplitude propagation of the form
$u_0(\xi)=u_0\exp[i(\lambda_u \xi+\phi_u)]$ and
$v_0(\xi)=v_0\exp[i(\lambda_v \xi+\phi_v)]$, where $\lambda_i$ and
$\phi_i$ correspond to the respective propagation constants and
initial phases, respectively, we obtain the following expression
for the power transfer: $\partial P_u/\partial \xi\sim \sin(\Delta \lambda \xi
+\Delta \phi)$, where $\Delta \lambda=\lambda_v-\lambda_u$ and $\Delta
\phi=\phi_v-\phi_u$. By assuming only a linear and local
dependence of the propagation constants, from (\ref{eqf}) we get:
$\lambda_u \approx C_0-\Delta k, \lambda_v \approx C_0/s_1$, which
results in $\Delta \lambda \approx \Delta k$. If the components
are initially in phase ($\Delta \phi=0$), the power transfer will
be initially towards the TE mode provided that $\Delta k>0$, and
towards the TM mode otherwise \cite{18}.

%
\begin{figure}[htb]
\scalebox{1}{\includegraphics{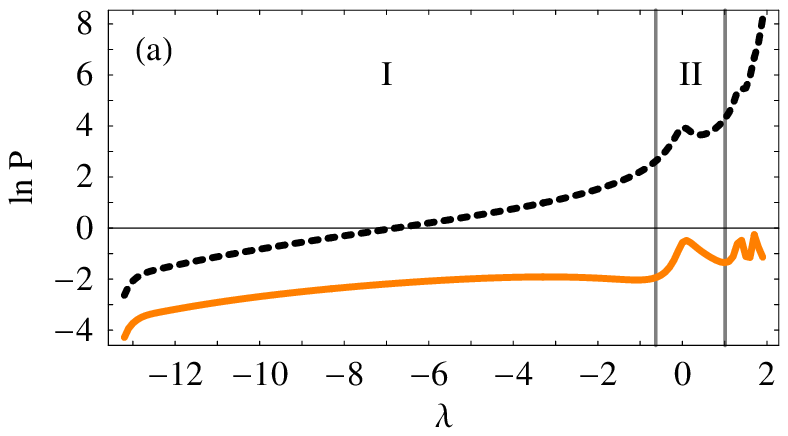}} \hfill
\scalebox{1}{\includegraphics{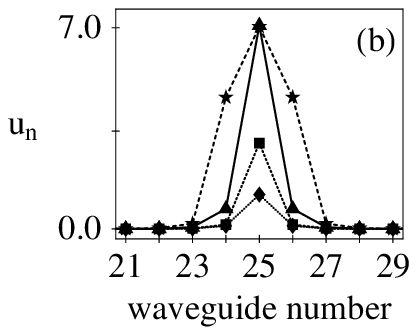}} \hfill
\scalebox{1}{\includegraphics{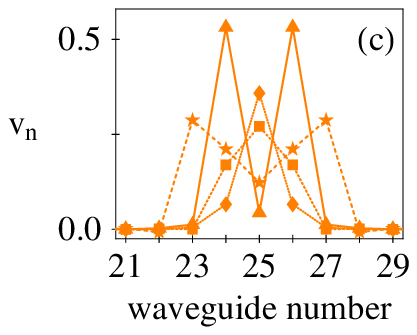}} \caption{(Color online).
(a) TE (dashed line) and TM (continuous line) powers as a function
of the propagation constant. Vertical lines separate regions. (b)
and (c) TE and TM profiles, respectively: Diamonds, squares,
triangles and stars correspond to $\lambda=-5,-1,0$, and $1.1$,
respectively. $C_0=\Delta k=\beta=10,\
V_0=1$.}\label{fig1}\vspace*{-3mm}
\end{figure}

To gain a theoretical background of the model (\ref{eqf}), we use
a Newton-Raphson method to find coupled localized stationary
solutions of the form: $u_n(\xi)=u_n\exp(i\lambda_u\xi)$ and
$v_n(\xi)=v_n\exp(i\lambda_v\xi)$, with $u_n,v_n\in\Re$. For the
sake of simplicity we assume $\lambda_u=\lambda_v\equiv\lambda$,
which in turn disables the power exchange between DVS components
[$\mathrm{Im}(v_n u_n^*)=0$ in (\ref{Pes})]. These solutions may
be regarded as the final stage of mode profiles after the DVS is
formed. The power dependence of the two components on propagation
constant, for the chosen set of experimentally achievable
parameters \cite{27}, is shown in Fig.~\ref{fig1}(a) in a
logarithmic scale. The region of existence of localized modes is
between the low-amplitude and high-amplitude limits for the upper
band edge plane wave~\cite{24} of the composed system of equations
(\ref{eqf}). In the present case, this region corresponds to
$\sim\lambda\in\{-13,2\}$. Power of the TE mode always exceeds
that of TM polarization and, interestingly, grows in a similar
fashion as the power of the on-site mode A in Ref.~\cite{23}. For
any $\lambda$, the TE mode is always a one-hump structure [see
Fig.~\ref{fig1}(b)] which spreads transversally in the region of
saturation \cite{23}. We may separate the $P\lambda$ diagram in
smaller regions depending on the shape of the TM mode. In region I
[$\sim\lambda\in\{-13,-0.6\}$], the TM mode corresponds to a
one-hump structure [diamonds and squares in Fig.~\ref{fig1}(c)].
In region II [$\sim\lambda\in\{-0.6,1\}$], the TM mode corresponds
to a two-hump structure separated by only one site [triangles in
Fig.~\ref{fig1}(c)]. In the next regions, the TM mode increases
its distance between the two humps in an odd number of waveguides
[as an example, see stars in Fig.~\ref{fig1}(c)].

While the total power increases, local saturation takes place
\cite{24}. As $\Delta k>0$, the TE mode is the one which gains
power. Therefore the TE mode starts to increase its power locally
together with the TM mode [region I in Fig.~\ref{fig1}(a),
diamonds and squares in Fig.~\ref{fig1}(b,c)]. However, above some
critical level of power [region II in Fig.~\ref{fig1}(a),
triangles in Fig.~\ref{fig1}(b,c)] the local power in the center
site is too high and the only possibility for the TM mode to exist
is by exploring the neighborhood looking for a more stable
configuration. Then, the TE mode further increases its power but
now, due to saturation, by increasing the amplitudes in the next
sites [see stars in Fig.~\ref{fig1}(b)]. Again, the TM mode finds
a new configuration which is initially stable, but now the
separation between peaks consists of three sites [stars in
Fig.~\ref{fig1}(c)]. If we continue increasing the power we
observe that the TE mode preserves its one-hump structure, by
increasing its width, while the TM mode has a two-hump structure
where the separation between peaks continuously increases.
Therefore, the DVS is mostly TE polarized, except at tails which
have a dominating TM polarization. The linear stability analysis
of solutions coincides with the Vakhitov-Kolokolov
criterion~\cite{book}: modes are stable for $\partial P/\partial
\lambda>0$, and unstable otherwise. This implies that in region I
solutions are always stable and, in the next regions, there exist
both stable and unstable sub-regions [see Fig.~\ref{fig1}(a)].

\begin{figure}[t]
\includegraphics[width=7cm]{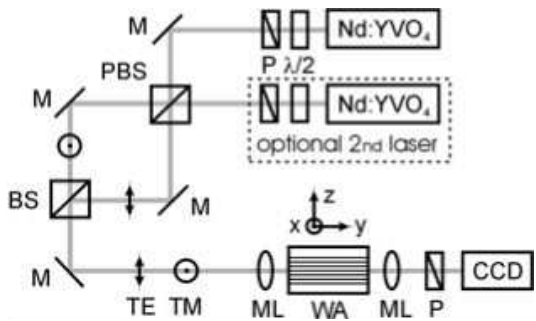}
\caption{Scheme of the experimental setup.}\label{fig2}\vspace*{-3mm}
\end{figure}

To verify our theoretical predictions we use the experimental
setup sketched in Fig.~\ref{fig2}. A cw laser with wavelength
532\,nm is split into two orthogonally polarized (TE and TM)
mutually coherent waves with the help of a polarizing beam
splitter PBS. Optionally, to allow for mutually incoherent
interaction of the two components, a TM polarized wave can be
provided by a second laser of the same wavelength. Input power is
adjusted with a combination of half wave plate $\lambda$/2 and
polarizer P. The two input beams are used to excite narrow
single-channel TE and TM polarized modes of the WA by using a
$40\times$ microscope lens ML. This nonlinear WA is fabricated in
x-cut lithium niobate doped with copper. The length of our sample
along the propagation y-direction is 11\,mm and the array consists
of 250 parallel titanium in-diffused waveguide channels that are
4\,$\mu$m wide with a separation of 4.4\,$\mu$m (grating period
$\Lambda=8.4\,\mu$m) \cite{28}. A second microscope lens ML images
the intensity on the output face onto a CCD camera, where an
additional polarizer P allows for independent observation of both
TE and TM components of the DVS.

\begin{figure}[t]
\includegraphics[width=8cm]{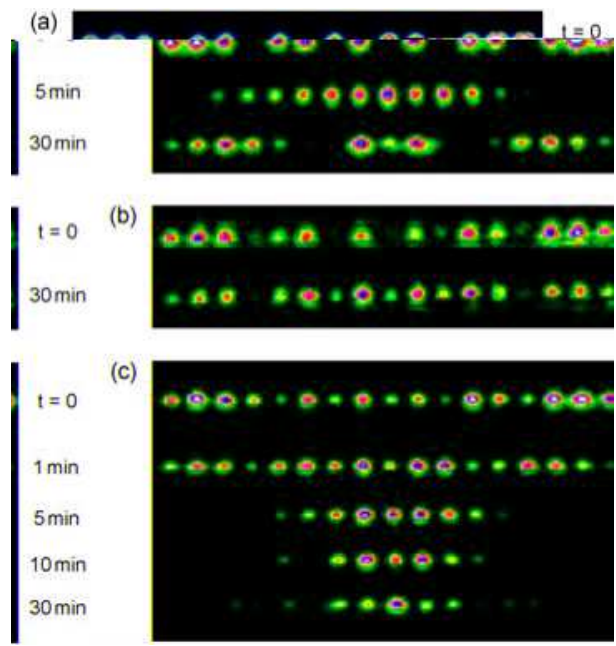}
\caption{(Color online). Experimentally observed DVS for mutually
coherent input beams. Temporal nonlinear evolution of (a) TE
component alone ($P_u=150\,\mu$W); (b) TM component alone
($P_v=300\,\mu$W); and (c) both components together
($P_u=150\,\mu$W, P$_v=300\,\mu$W).}\label{fig3}\vspace*{-3mm}
\end{figure}

The nonlinear dynamics of TE only, TM only, and both TE and TM
modes (mutually coherent from the same light source) is presented
in Fig.~3. Here we make use of the fact that in photorefractives
the nonlinearity grows exponentially in time,
$\beta(t)=\beta\,(1-\exp[-t/\tau])$, where $\tau$ is the
dielectric response time \cite{30}. Initially, after switching on
the light ($t=0$), discrete diffraction is monitored for each
situation. Although one may observe initial focusing of the TE
mode within the first minutes in Fig.~\ref{fig3}(a) (an even weaker effect
is observed for TM), both modes alone are incapable to form a
localized structure [Fig.~\ref{fig3}(a,b)]. However, when both input
polarizations are present [Fig.~\ref{fig3}(c)] a five-channel wide DVS is
formed after $t=30$\,min and remains stable for longer times $t$.

\begin{figure}[h]
\includegraphics[width=6.5cm]{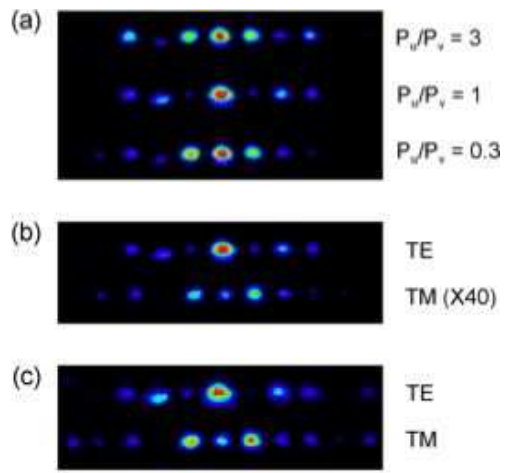}
\caption{(Color online). Mode analysis of stationary DVS solitons.
(a) Stationary output (total intensity TE + TM) of DVS for
different power ratios $P_u/P_v$; (b) stationary TE and TM
(amplified 40$\times$) polarized components for $P_u/P_v=1$; and
(c) stationary TE and TM polarized components for $P_u/P_v=1$ when
input TE beam is blocked after formation of the DVS.}\label{fig4}
\end{figure}

Stationary images of DVS collected from the output facet of the
sample for a fixed value of TE power and different values of TM
power are presented in Fig.~\ref{fig4}(a). As can be seen, the shape of the
DVS slightly changes for different power ratios $P_u/P_v$. TE and
TM polarized components for a power ratio $P_u/P_v=1$ (steady
state) are shown in Fig.~\ref{fig4}(b). As predicted, a dominating
single-hump TE polarized component and a weaker double-humped TM
component are observed. The role of the TM input polarization can
be further analyzed by blocking the TE input after stable
formation of the DVS in Fig.~\ref{fig4}(c). Obviously the TM polarized
input light transfers most of its energy to the TE component,
forming a single-hump solution, while the remaining power is
trapped in form of a two-hump solution. This energy transfer, from
ordinary to extraordinary polarization (TM $\rightarrow$ TE), is
due to a specific anisotropic nonlinearity in LiNbO$_{3}$ [in
model (\ref{eqf}), this corresponds to consider $\Delta
k>0$]. The mechanism of coupling of orthogonally polarized modes
is explained by writing holographic gratings due to photovoltaic
currents. Light is anisotropically diffracted from these shifted
gratings with polarization conversion, which leads to an energy
exchange among the modes \cite{29}.

\begin{figure}[h]
\includegraphics[width=7cm]{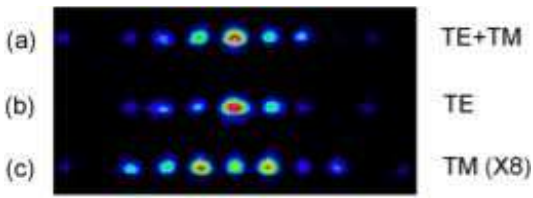}
\caption{(Color online). Experimentally observed DVS for mutually
incoherent input beams. (a) Stationary output of DVS [total
intensity TE + TM] for $P_u/P_v=1.5$; (b) TE polarized component;
and (c) TM polarized components (amplified
8$\times$).}\label{fig5}
\end{figure}

Energy coupling of orthogonally polarized waves can be prevented
by using mutually incoherent input beams ($B=0$). Experimentally
this is realized by a second laser of the same wavelength (see
Fig.~\ref{fig2}), which now provides the TM polarized input beam,
and corresponding  results for the steady-state DVS formation are
shown in Fig.~\ref{fig5}. Again a two-hump structure is observed
for the TM component, which now guides a significant part of the
total power of the soliton. 

In conclusion, we suggested a rather general theoretical model to
describe saturable discrete vector solitons having orthogonally
polarized components. Power transfer and coupling between TE and
TM components is investigated as well as the corresponding
localized stationary solutions. We discovered that these composite
solitons might have different width and shape depending on the
region of parameters. The dominating TE mode is single-humped
while the weaker TM mode may exhibit both one- and two-hump
structures. We confirm our findings experimentally by using either
coherent or mutually incoherent excitations, where the latter is
used to suppress energy coupling in formation of discrete vector
solitons. Our experimental conditions match the
region II of stationary solutions, a region with an intermediate
level of power and highly localized solutions. This is because a
one-channel input excites a strongly localized region of the array
with high local intensity. The results obtained here could be useful in the
codification of signals, filtered by polarization, in future
all-optical communication networks.

We gratefully acknowledge financial support from BMBF (grant
DIP-E6.1), DFG (grant KI482/8-1), and MNZ\v ZSRS (grant 14-1034).

\end{document}